\documentclass[12pt]{article}
\usepackage{graphicx}
\usepackage{amsfonts}
\usepackage{amssymb}
\newcommand{\nin}{\noindent}
\newcommand{\be}{\begin{equation}}
\newcommand{\ee}{\end{equation}}
\newcommand{\beq}{\begin{eqnarray}}
\newcommand{\eeq}{\end{eqnarray}}

\newcommand{\hf}{\frac{1}{2}}
\newcommand{\nn}{\nonumber\\}

\newcommand{\di}{\partial_i}
\newcommand{\djj}{\partial_j}
\newcommand{\dk}{\partial_k}
\newcommand{\dl}{\partial_l}
\newcommand{\dt}{\partial_t}
\newcommand{\dmu}{\partial_\mu}
\newcommand{\dnu}{\partial_\nu}
\newcommand{\e}{\vec{ e}}
\newcommand{\ex}{\vec{ e}(x)}
\newcommand{\die}{\di \vec{ e}}
\newcommand{\dje}{\djj \vec{ e}}
\newcommand{\dke}{\dk \vec{ e}}
\newcommand{\dle}{\dl \vec{ e}}
\newcommand{\dte}{\dt \vec{ e}}

\newcommand{\diedie}{\die \cdot \die}
\newcommand{\diedje}{\die \cdot \dje}
\newcommand{\djedje}{\dje \cdot \dje}
\newcommand{\ebe}{\e \cdot \triangle \e}
\newcommand{\pn}{\vec{\pi}}
\newcommand{\pni}{\di \pn}
\newcommand{\pnj}{\djj \pn}

\newcommand{\pnt}{\dt \pn}
\newcommand{\pipnt}{\pn \cdot \dt \pn}
\newcommand{\dith}{\di \theta}
\newcommand{\dithh}{\dith \dith}
\newcommand{\dtth}{\dt \theta}
\newcommand{\boxth}{\triangle\theta}
\newcommand{\kp}{\vec{ k}}
\newcommand{\pp}{\vec{ p}}
\newcommand{\intom}{\int \frac{d\omega}{2 \pi}}
\newcommand{\intka}{\int \frac{d^2 k}{(2 \pi)^2}}
\newcommand{\prop}{\omega^2 - k^4 + i \epsilon}
\newcommand{\Jd}{\tilde{J}}

\begin{document}

\begin{center}

{\Large{\bf  Non-Linear Sigma Model and asymptotic freedom at the Lifshitz
point}}

\vspace{0.5cm}

K. Anagnostopoulos$^a$\footnote{konstant@mail.ntua.gr}, K. Farakos$^a$\footnote{kfarakos@central.ntua.gr}, P.
Pasipoularides$^a$\footnote{p.pasip@gmail.com}, A. Tsapalis $^{a,b}$\footnote{a.tsapalis@iasa.gr}

\vspace{0.5cm}

$^a$ Department of Physics, National Technical University of
       Athens \\ Zografou Campus, 157 80 Athens, Greece

$^b$ Hellenic Naval Academy, Hatzikyriakou Avenue, Pireaus 185 39, Greece

\vspace{2cm}

{\bf Abstract}

\end{center}

\nin

We construct the general $O(N)$-symmetric non-linear sigma model in
$2+1$ spacetime dimensions at the Lifshitz point with dynamical critical
exponent $z=2$. For a particular choice of the free parameters, the model is
asymptotically free with the beta function coinciding
to the one for the conventional sigma model in 1+1 dimensions.
In this case, the model admits also a simple description
in terms of adjoint currents.

\eject

\section{Introduction}

Quantum field theories in the Lifshitz context have received a considerable
amount of investigation recently as their renormalizability properties
are radically altered compared to the conventional Lorentz symmetric
theories~\cite{field}.
A Lifshitz type theory is based on the anisotropic behavior between
spatial and temporal directions under scale transformations:
$t \to b^z t$ and ${\bf x} \to b~ {\bf x}$ where the degree of anisotropy is
measured by the dynamical critical exponent $z$. As a result, plane waves
propagate with dispersion relation $\omega = p^z$ in this theory, where
$\omega$ denotes energy and $p$ is the magnitude of spatial momentum.
In the quantized theory, higher power of momenta appear in the
denominator of the free field propagator
$ ~i (\omega^2 - p^{2z} + i \epsilon)^{-1}$ which lower the superficial degree
of divergence of perturbative graphs, and render new operators
renormalizable at the Lifshitz ultraviolet (UV) point. 
Renormalizable theories of gravity in 3+1
dimensions have been proposed by Ho\v{r}ava~\cite{Horava1, Horava2}
and triggered a large amount of
subsequent investigations on the nature of the flow to conventional
general relativity in the low energy regime as well as on
cosmological or black hole solutions, see for example ~\cite{lifshitz-cosmology, lifshitz-blackholes} and references there in. In addition, a relation 
between the phase diagram in Ho\v{r}ava-Lifshitz gravity and 
causal dynamical triangulations quantum gravity is found 
in~\cite{Ambjorn:2010hu}.

The renormalization of various field theoretical models in flat spacetime at
the Lifshitz point has been examined
already. Ho\v{r}ava formulated Yang-Mills (YM) theory in 4+1 dimensions at the
Lifshitz point with $z=2$ and
showed that the dimensionless coupling is asymptotically
free~\cite{HoravaYM}.
Electrodynamics has also been formulated at the Lifshitz point \cite{QED}.
The $CP^{N-1}$ model has been constructed in $2+1$ dimensions
with $z=2$ and a large-$N$ analysis has shown that the model is
asymptotically free~\cite{CPN}.
In addition, the Liouville theory becomes renormalizable in
3+1-d Lifshitz spacetime with $z=3$~\cite{LL},
a fact that may assist further cosmological
investigations of Ho\v{r}ava's gravity.
The four-fermion interaction is renormalizable in
3+1 dimensions with $z=3$~\cite{4Fermi}
while divergences of the Standard Model (SM)
interactions become softer~\cite{SM}
as for example in the Yukawa model~\cite{Yukawa},
where only logarithmic divergences remain. Such behavior is promising
in dealing with the hierarchy of masses in the SM.

From the Lifshitz UV perspective the Lorentz-symmetric gaussian terms
become relevant operators and are generically expected
to dominate the infrared (IR)
regime of the theory. Even if absent from the classical action, quantum
corrections will generate such terms which approximately
restore Lorentz symmetry in the low energy effective action as has been
demonstrated in the Yukawa~\cite{Yukawa} and Liouville~\cite{LL} theory.
Note however that if the model contains more than one species of
interacting particles, the recovery of the speed of light for all the modes
requires the fine-tuning of bare parameters~\cite{iengo}.

A particular class of anisotropic actions at $z=2$ are the
so-called {\it detailed balance} actions. These actions are constructed
in $D+1$-dimensional spacetime from the squaring of the equations of motion
of the Euclideanized $D$-dimensional action $W[\phi]$, i.e. the spatial
part of the Lifshitz Lagrangian is proportional to
\be
\left( \frac{\delta W[\phi]}{\delta \phi} \right)^2 ~.
\label{db}
\ee
The relation of these actions to the stochastic quantization scheme is  
discussed in~\cite{Dijkgraaf:2009gr}.
There is evidence that the detailed balance action
with the potential term~(\ref{db}) inherits the quantum
properties of the $D$-dimensional theory $W[\phi]$,
in the sense that the RG flow of marginal couplings in both theories
is the same. This is indeed the case in the 4+1-d YM theory constructed
by Ho\v{r}ava~\cite{HoravaYM} which has an asymptotically free coupling
with the well known beta function of standard YM in four dimensions.
This is in
sharp contrast to the conventional YM in five dimensions which is
non-renormalizable. We note that discretizations of conventional YM
{\it ala} Wilson on 5-d Euclidean
lattices lack a continuum limit (in the sense that the order-disorder
phase transition is not continuous)
unless anisotropic couplings in the spatial
and temporal directions are introduced~\cite{farakos}.
On the other hand,
Ho\v{r}ava's model if properly discretized
will possess the continuum limit and in that sense provides
the UV completion of gauge theory in five dimensions.
As a warm-up exercise we investigate the detailed balance action for the
scalar theory in five dimensions. We utilize the general
renormalization group (RG) flow study
of~\cite{iengo} and show in Appendix A that the beta function for the
marginal coupling of the 5-d detailed balance action is identical to
the beta function of the standard 4-d theory.

The aim of this paper is to investigate the issue of quantum inheritance due
to the spacetime anisotropy in the context of the
$O(N)$-symmetric non-linear sigma model (NLSM) in 2+1 dimensions.
It is well known that the $O(N)$ NLSM is asymptotically free in
1+1 dimensions and non-renormalizable in higher dimensions.
It is reasonable therefore to expect that an asymptotically free
NLSM exists in three dimensions if spacetime becomes anisotropic.
We construct the general NLSM in 2+1 dimensions at the $z=2$ Lifshitz point
through the identification of all the marginal and relevant
operators allowed by symmetry. We analyze perturbatively the model at one
loop and identify the one-coupling model
which possesses asymptotic freedom. It turns out that this 'tuned' action
shares a common beta function with the standard NLSM in 1+1 dimensions.
In addition, the Lorentz-symmetric relevant
operator does not appear in the effective action.
Low energy pions will therefore propagate
with a non-relativistic dispersion relation $\omega^2 = p^4 + m^4$.

The structure of the paper is as follows: In Section 2 we present a brief
review of the NLSM in 1+1 dimensions.
In addition, we present an
equivalent formulation of the theory in terms of adjoint currents.
 Section 3 deals with the NLSM in 2+1 dimensions at the $z=2$ Lifshitz
point. The construction of the general action is presented in
Section 3.1.
In Section 3.2 we perform a perturbative
analysis of the model at one loop.
In the last subsection (3.2.2) we formulate the asymptotically free model in
terms of the adjoint currents. Section 4 contains the conclusions of this
study. In Appendix A.1 we review some standard properties of scalar
field theory and in Appendix A.2 we examine the quantum inheritance
property for the scalar theory between four euclidean and five
anisotropic dimensions.
Finally, in  Appendix B we examine the abelian rotor -or XY model- at the
Lifshitz point.

\section{1+1-Dimensional Non Linear Sigma Model}

The $O(N)$ invariant NLSM is defined in 1+1 spacetime dimensions
through a multiplet of scalar fields $\e$ which obey the unimodulus
constraint at each spacetime point $x$:
\begin{equation}
\vec{e}(x)=\left( e_0(x),e_1(x),..,e_{N-1}(x)\right) ~~~,~~~~~
\vec{e}(x)\cdot \vec{e}(x)=1
\end{equation}
with action
\be
\label{W2}
W[\e] = \frac{1}{2g^2} \int d^2 x \; \diedie ~.
\ee
The quantization of the model is performed by the functional integration
\be
\label{Z2}
Z[\e] = \int D\e \prod_x \delta \left(\ex^2-1 \right) e^{-W[\e]}~.
\ee
The starting point for a perturbative study
of the general $O(N)$
action~(\ref{W2}) is the introduction of a constrained field $\sigma$,
and $N-1$ pion fields $\pn$,
\be
\e(x) = \left(\sigma(x) \;,g ~\pn (x)\right), ~~\pn (x)=\left (\pi_1(x),\pi_2(x),..,\pi_{N-1}(x)\right)
\ee
such that
\be
\sigma = \sqrt{1-g^2 \pn^2} = 1 -\hf g^2 \pn^2 + {\cal O}(g^4)~.
\ee
Although the expansion in $g$ generates infinite pion vertices --the lowest
interaction is the four-pion vertex at ${\cal O}(g^2)$--, the theory is
renormalizable to all-orders in perturbation theory~\cite{brezin}.
An ${\cal O}(g^2)$ evaluation of the two-point function determines
easily the beta function of the model, (e.g.~\cite{Peskin}),
\be
\beta(g) = -\frac{N-2}{4 \pi}g^3
\ee
which is {\it asymptotically free}.

Although classically the model possesses a continuum of vacua belonging in the
$O(N)/O(N-1)$ coset space, the excitations above the vacua do not remain
massless in the quantum theory. This is a demonstration of the
Coleman-Mermin-Wagner theorem~\cite{Coleman}
which states that a continuous symmetry cannot
break spontaneously in two dimensions (at finite temperature --or equivalently
finite values of the coupling $g$) and has its origin in the infrared
singularities of the theory. In other words, quantum fluctuations disorder
the system and a mass gap is generated dynamically.

Let us also note an equivalent formulation of the model which will become
relevant in the construction of the Lifshitz model.
The 2-d NLSM action~(\ref{W2}) can be expressed in terms of the
$N(N-1)/2$ conserved currents $J_{\mu}^{(a,b)}$ of the theory
\be
J_{\mu}^{(a,b)} = e^a \partial_{\mu} e^b - e^b \partial_{\mu} e^a
~~~~~~~~(a,b = 1,...,N)
\ee
as
\be
W[\e] = \frac{1}{4g^2} \int d^2 x J_{\mu}^{(a,b)}J_{\mu}^{(a,b)} =
- \frac{1}{4g^2} \int d^2 x \,tr {\bf J_{\mu}} {\bf J_{\mu}}
\ee
with ${\bf J_{\mu}}$ viewed also as $N \times N$ matrix transforming in the
adjoint representation of the internal space.
The equation of motion for the fields is then expressed as the
conservation of the current $J_{\mu}$
\be
\dmu {\bf J_{\mu}} = 0
\ee
Indeed since
\be
\dmu J_{\mu}^{(a,b)} = e^a \triangle e^b - e^b \triangle e^a
\label{dJ1}
\ee
and dotting with $e^b$ leads to
\be
\dmu J_{\mu}^{(a,b)} e^b = e^a (\ebe) - \triangle e^a = 0
\ee
the well known equations of motion for the $\e$ fields in the presence of
the unimodulus constraint are reproduced.

Note finally that a dual current can also be defined in two dimensions as
\be
{\bf \Jd_\mu} = \epsilon_{\mu\nu} {\bf J_\nu}
\ee
which  is {\it not conserved} since
\be
\dmu \Jd_{\mu}^{(a,b)} = 2 \epsilon_{\mu \nu}
\dmu e^a \partial_\nu e^b ~.
\label{dJ2}
\ee

\section{The Non Linear Sigma Model at the $z=2$ Lifshitz point}

\subsection{The general NLSM}

In this section we construct the $O(N)$-symmetric non-linear sigma
in $2+1$ spacetime dimensions
at the Lifshitz-type fixed point in the UV,
with dynamical critical exponent $z=2$.

The Lagrangian ${\cal L}$ of the model should respect $O(N)$-symmetry,
and consists of the kinetic term ${\cal L}_K$ and a
potential term $ {\cal L}_V$.
\begin{equation}
{\cal L}={\cal L}_K - {\cal L}_V
\end{equation}
The kinetic  term is of the form:
\begin{equation}
{\cal L}_K=\frac{1}{2 g^2}\partial_t \vec{e}\cdot \partial_t\vec{e}
\end{equation}
where $g$ is a dimensionless coupling ($[g]=0$), while the
canonical dimensions of $t$, $x$, and $\vec{e}$ are
\begin{equation}
[t]=-2, ~~~ [x]=-1, ~~~[\vec{e}]=0 ~.
\end{equation}
The potential term includes all the marginal (with dimension $D+z=4$)
and relevant $O(N)$-symmetric operators.
There exist three marginal
\footnote{~~Due to the unimodulus constraint, $\e \cdot \die = 0$,
$\die \cdot \die = -\e \cdot \triangle \e$, and therefore the operator
${\cal{O}}_{2}$ is equivalent to $(\ebe)^2$.
Integrating by parts, ${\cal{O}}_{3}$ is equivalent to
$(\e \cdot \di \djj \e)(\e \cdot \di \djj \e) $.
}
\footnote{$~~\di$ denotes now
a spatial derivative and $\triangle = \di \di$. }
\beq
{\cal{O}}_{1} &=& \triangle \e \cdot \triangle \e  \nn
{\cal{O}}_{2} &=& \left(\diedie \right)^2     \nn
{\cal{O}}_{3} &=& \left(\diedje \right)\left(\diedje \right) ~,
\eeq
and one relevant, dimension two, operator
\beq
{\cal{O}}_R= \partial_{i} \vec{e} \cdot \partial_{i} \vec{e} ~.
\eeq
The general potential term is hence of the form:
\beq
{\cal L}_V=\frac{1}{2 g^2}
\left(\eta_1 {\cal{O}}_{1}+\eta_2 {\cal{O}}_{2}+\eta_3 {\cal{O}}_{3}
 +M^2 \cal{O}_R\right)
\eeq
where the couplings $\eta_1, \eta_2$ and $\eta_3$ are dimensionless.
The most general renormalizable $O(N)$ action at the $z=2$ Lifshitz point
is therefore
\beq
\label{Son}
S_{z=2}[\e] = \frac{1}{2g^2} \int dt d^2 x \;
\Big[ \dte \cdot \dte
-\triangle \e \cdot \triangle \e
- \eta_2 \left(\diedie \right)^2 \nn
- \eta_3 \left(\diedje \right)\left(\diedje \right)
- M^2 \diedie\Big]
\eeq
Note that the last term (if present)
will restore the Lorentz symmetry in the low energy
$|\vec{k}| \ll M$ regime. The coupling $\eta_1$ is redundant as it can be
set to 1 without loss of generality by a suitable
rescaling of space and time coordinates.

\subsection{The asymptotically free model}

The general action~(\ref{Son}) constructed in the previous sections
contains three dimensionless couplings.
The flow of the couplings in general requires the examination of
four-point functions and goes beyond the scope of this work. In contrast,
our aim is to investigate the existence of asymptotic freedom in $2+1$
spacetime dimensions for the Lifshitz sigma model.
Therefore, keeping in mind the Lorentzian 2-d model,
we will examine the flow of the dimensionless coupling $g$
considering the other two couplings $\eta_2, \eta_3$ as {\it fixed
multiplicative constants}.

\subsubsection{Perturbative analysis }

Following the standard analysis of the 2-d model (e.g.~\cite{Peskin}) we will
examine perturbatively the two-point function of the model at 1-loop.
Solving the unimodulus constraint in terms of
$\sigma = \sqrt{1-g^2 \pn^2}$ and $N-1$ pions $\vec{\pi}$ we have
\be
\sigma = \sqrt{1-g^2 \pn^2} = 1 -\hf g^2 \pn^2 + O(g^4)
\ee
and we obtain the expressions at $O(g^4)$
\be
\partial_{i}\sigma=-g^2 \left(\vec{\pi }\cdot \partial_{i}\vec{\pi}\right),~~~ \triangle \sigma=-g^2 \left((\partial_{i}\vec{\pi})^2+\vec{\pi }\cdot \triangle \vec{\pi}\right)~.
\ee
The $O(g^2)$ pion action therefore which is amenable to the perturbative
treatment is written
\beq
\label{Spn}
&&S_{z=2}[\pn] = \frac{1}{2} \int dt d^2 x \;
\Big[ \pnt \cdot \pnt
- \triangle\pn \cdot \triangle \pn + g^2 (\pipnt)^2 \\
&&- g^2 \big[ (\pni)^2 + \pn \cdot \triangle \pn \big]^2
- \eta_2 g^2 (\pni \cdot \pni)^2
-\eta_3 g^2 (\pni \cdot \pnj)(\pni \cdot \pnj)
\Big] \nonumber
\eeq
where the relevant operator -proportional to $M^2$- has been omitted as it
will not affect the UV divergences of the theory.

\vspace{0cm}

\begin{figure}[ht]
\begin{center}
\includegraphics[width=12cm]{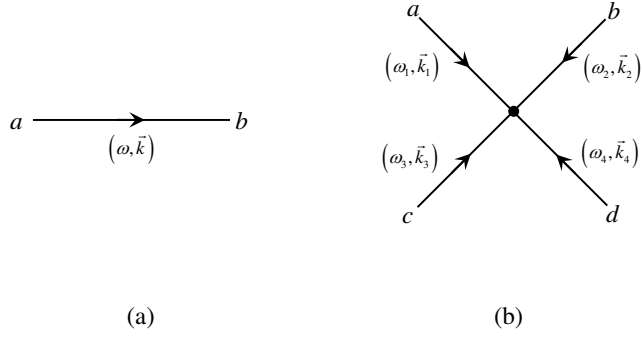}
\caption{Feynman rules for the Lifshitz NLSM.
(a) the $z=2$ Lifshitz propagator, eq.~(\ref{prop}).
(b) the four-pion vertex, eqs.~(\ref{v1}, \ref{v2}, \ref{v3}, \ref{v4}).}
\end{center}
\end{figure}

The bare pion propagator has the form
\be
\label{prop}
G^{ab}(\omega,\vec{k}) =
\left< \pi^a(-\omega,-\vec{k}) \pi^b(\omega,\vec{k}) \right>
=\frac{i}{\omega^2- k^4 + i \epsilon} \delta^{ab}
\ee
where $k = |\vec{k}|$.
From~(\ref{Spn}) we deduce the following Feynman rules for the $\pi^4$
interactions symmetrizing appropriately the vertex.
For example the first $O(g^2)$ term gives
\beq
i\frac{g^2}{2}(i\omega_2)(i\omega_4) \delta^{ab}\delta^{cd} \rightarrow
\frac{-i g^2}{24} \Big[
(\omega_1 + \omega_2)(\omega_3 + \omega_4) \delta^{ab}\delta^{cd} + \nn
(\omega_1 + \omega_3)(\omega_2 + \omega_4) \delta^{ac}\delta^{bd} +
(\omega_1 + \omega_4)(\omega_2 + \omega_3) \delta^{ad}\delta^{bc} \Big]
\label{v1}
\eeq
\nin
The second vertex is
\beq
&&-i\frac{ g^2}{2} \Big[(i\kp_1 \cdot i\kp_2) (i\kp_3 \cdot i\kp_4)
\delta^{ab}\delta^{cd} +
 \kp_2^2 \;\kp_4^2 \, \delta^{ab}\delta^{cd} -
2 (i\kp_1 \cdot i\kp_2)\kp^2_4 \, \delta^{ab}\delta^{cd}
\Big] \rightarrow \nn
&& \frac{-i g^2}{24} \Big[
4 (\kp_1 \cdot \kp_2) (\kp_3 \cdot \kp_4)+
(\kp_1^2 + \kp_2^2)(\kp_3^2 + \kp_4^2) +
2 (\kp_1 \cdot \kp_2)(\kp_3^2 + \kp_4^2) + \nn
&&\hspace{2cm}
2 (\kp_1^2 + \kp_2^2)(\kp_3 \cdot \kp_4) \Big]\, \delta^{ab}\delta^{cd}
+ ({\rm two \,\, cycl. \,\, perms.})
\eeq
which nicely simplifies to
\beq
\frac{- i  g^2}{24} \Big[
(\kp_1 + \kp_2)^2(\kp_3 + \kp_4)^2  \delta^{ab}\delta^{cd} +
(\kp_1 + \kp_3)^2(\kp_2 + \kp_4)^2  \delta^{ac}\delta^{bd}  \nn
+(\kp_1 + \kp_4)^2(\kp_2 + \kp_3)^2  \delta^{ad}\delta^{bc}
\Big]
\label{v2}
\eeq

\nin
The third vertex will be
\beq
-i\frac{\eta_2 g^2}{2}(i\kp_1 \cdot i\kp_2) (i\kp_3 \cdot i\kp_4)
\delta^{ab}\delta^{cd}
\rightarrow
\frac{-i \eta_2 g^2}{6} \Big[
(\kp_1 \cdot \kp_2) (\kp_3 \cdot \kp_4)
\delta^{ab}\delta^{cd} + \nn
(\kp_1 \cdot \kp_3) (\kp_2 \cdot \kp_4)
\delta^{ac}\delta^{bd} +
(\kp_1 \cdot \kp_4) (\kp_2 \cdot \kp_3)
\delta^{ad}\delta^{bc} \Big]
\label{v3}
\eeq
\nin
Finally the last interaction term produces a vertex
\beq
-i\frac{\eta_3 g^2}{2}(i\kp_1 \cdot i\kp_3) (i\kp_2 \cdot i\kp_4)
\delta^{ab}\delta^{cd}
\rightarrow
\frac{- i  \eta_3 g^2}{12} \Big[
(\kp_1 \cdot \kp_3) (\kp_2 \cdot \kp_4)
(\delta^{ab}\delta^{cd} + \delta^{ad}\delta^{bc} ) \nn
+ (\kp_1 \cdot \kp_2) (\kp_3 \cdot \kp_4)
(\delta^{ac}\delta^{bd} + \delta^{ad}\delta^{bc} )
+ (\kp_1 \cdot \kp_4) (\kp_2 \cdot \kp_3)
(\delta^{ab}\delta^{cd} + \delta^{ac}\delta^{bd} )
\Big] \nn
\label{v4}
\eeq

\vspace{0cm}

\begin{figure}[ht]
\begin{center}
\includegraphics[width=12cm]{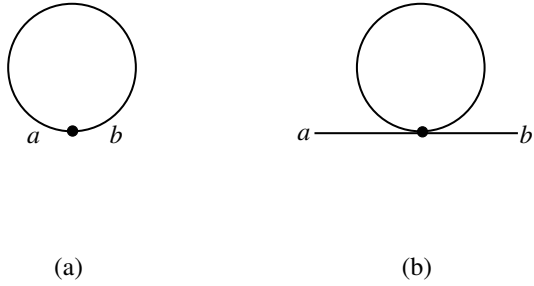}
\caption{The one-loop diagrams relevant to the renormalization of
the model. (a) the bubble relevant to the wave function renormalization $Z$
at ${\cal{O}}(g^2)$.
(b) one-loop contribution to the pion two-point function.}
\end{center}
\end{figure}

\nin
The one loop corrections to the two point function (Figure 2b)
can be computed easily
from the above vertices considering all possible contractions of two pion
fields. The symmetry factor for the two point function graph is 12.
Denoting $(\omega,\kp)$ the internal and $(p_0,\vec{p})$
the external energy/momentum we get from the first vertex a total contribution
\be
C_1 = - i g^2 \intom \intka i \frac{-(\omega + p_0)^2}{\prop}
\delta^{ab}
\ee

\nin
From the second vertex we get a contribution
\be
C_2 = - i g^2 \intom \intka i \frac{(\kp + \pp)^2 (\kp + \pp)^2}{\prop}
\delta^{ab}
\ee

\nin
From the third vertex we get a contribution
\be
C_3 = - i \eta_2 g^2 \intom \intka i \frac{2 (N-1)\; \kp^2\; \pp^{\;2}
+ 4  (\kp \cdot \pp)^2 } {\prop} \delta^{ab}
\ee
The last vertex contributes
\be
C_4 = - i \eta_3 g^2 \intom \intka i \frac{2\; \kp^2\; \pp^{\;2}
+ 2 N (\kp \cdot \pp)^2 } {\prop} \delta^{ab}
\ee

\nin
The total contribution $(C_1 + C_2 + C_3 + C_4) = C_{tot} \delta^{ab}$
is therefore (odd terms vanish)
\be
C_{tot} = g^2 \intom \intka \frac{-\omega^2 - p_0^2 + k^4 + p^4 +
[4 + 2 N \eta_2 + (N+2) \eta_3] k^2 p^2 }
{\prop}
\ee
where the two-dimensional symmetric integration property has been used
\be
\intka (\kp \cdot \pp)^2 f(k^2) = \frac{1}{2} \intka k^2 p^2 f(k^2)
\ee
After a Wick rotation the $\omega$ integration is performed easily
picking up poles at $\omega = \pm k^2$ through the usual Feynman prescription.
The result is
\beq
C_{tot} =&& -i g^2 \delta^3 (0) + i (p_0^2-p^4) g^2 \intka \frac{1}{k^2}
\nn
&&-i p^2 g^2  \Big[ 4 + 2 N \eta_2 + (N+2) \eta_3 \Big] \intka
\label{ctot}
\eeq
The first term in eq.~(\ref{ctot}) is an infinite constant that can be dropped.
The second term is precisely the Lifshitz free pion action which corresponds
to the propagator~(\ref{prop}) and renders the model renormalizable through
the logarithmically divergent integral. The third term implies the
generation of a $p^2$ dependent term in the effective action which diverges
with $\Lambda^2$ ($\Lambda$ is the momentum cutoff)
that was absent in
the bare marginal theory. In order to perturbatively renormalize the model
we will require the vanishing of the third term. This is possible only if
\be
4 + 2 N \eta_2 + (N+2) \eta_3 = 0
\ee
which in turn requires for generic $N$
\be
\eta_2 = 1 \hspace{1cm} \eta_3 = -2
\ee
For these particular values of the coefficients $\eta_2, \eta_3$ the
behavior of the divergences becomes identical to the
ordinary 2-d NLSM.
The model will be renormalizable at one-loop with the redefinition of the
scale dependent coupling $g$, and wave function renormalization constant
$Z$. The dependence on the scale $M$ is contained in
the $\beta(g)$ and $\gamma(g)$ functions:
\be
\beta(g) =  M\frac{\partial}{\partial M} g  ~~~~~,~~~~~
\gamma(g) =  M\frac{\partial}{\partial M} \log \sqrt{Z} ~.
\ee
For their extraction at leading order,
it is enough to consider the Callan-Symanzik equation for the
$\langle \sigma(0) \rangle$ and $\langle e_a(p) e_b(-p) \rangle$
($a,b=1,\dots N-1$) correlation functions:
\beq
&&\left( M\frac{\partial}{\partial M} + \beta(g) \frac{\partial}{\partial g} +
\gamma(g) \right) \langle \sigma(0) \rangle = 0
\label{CSZ1}\\
&&\left( M\frac{\partial}{\partial M} + \beta(g) \frac{\partial}{\partial g} +
2 \gamma(g) \right) \langle e_a(p_0,p)~e_b(-p_0,-p) \rangle = 0
\label{CSZ2}
\eeq
At ${\cal{O}}(g^2)$ the relevant diagrams which contribute
are shown in Figure 2. The evaluation of the bubble in Figure 2a
gives at the subtraction scale $M$:
\beq
\langle \sigma(0) \rangle &=& 1 - \frac{g^2}{2} \langle \pi^2 (0)\rangle =
1-\frac{g^2}{2} (N-1) \intom \intka \frac{i}{\prop} \nn
&=& 1-\frac{g^2}{2} (N-1) \intka \frac{1}{k^2} =
1 - \frac{g^2 (N-1)}{8 \pi} \log \frac{M^2}{\mu^2}
\label{cor1}
\eeq
where the infrared cutoff $\mu$ is also introduced. Similarly,
the tree level and one-loop terms (Figure 2b) contribute to the
two-point function:
\beq
&&\langle e_a(p_0,p)~ e_b(-p_0,-p)  \rangle =
g^2\langle \pi_a (p_0,p)~ \pi_b (-p_0,-p) \rangle = \nn
&&g^2 \left( \frac{i}{p_0^2-p^4}+ \frac{i}{p_0^2-p^4} C_{tot}
\frac{i}{p_0^2-p^4} \right) \delta^{ab}
=\frac{i}{p_0^2-p^4}
\left(g^2 - \frac{g^4}{4 \pi} \log \frac{M^2}{\mu^2} \right) \delta^{ab}
\nn
\label{cor2}
\eeq
Plugging the results~(\ref{cor1}) and (\ref{cor2}) to the Callan-Symanzik
equations~(\ref{CSZ1}), (\ref{CSZ2}), we immediately obtain the
leading order beta function
\be
\beta(g) = -\frac{N-2}{4 \pi}g^3 + {\cal{O}}(g^5) ~.
\ee
We conclude therefore that an asymptotically free NLSM exists in 2+1
spacetime dimensions at the $z=2$ Lifshitz point with action
\beq
S_{\mbox{\scriptsize asym.~free}}[\e] &=& \frac{1}{2g^2} \int dt d^2 x \;
\Big[ \dte \cdot \dte - {\cal{O}}_{1} - {\cal{O}}_{2}
+ 2~ {\cal{O}}_{3} \Big] \nn
&=& \frac{1}{2g^2} \int dt d^2 x \;
\Big[ \dte \cdot \dte
-\triangle \e \cdot \triangle \e
- \left(\diedie \right)^2 \nn
&& \hspace{3.8 cm} +~ 2~\left(\diedje \right)\left(\diedje \right)\Big]
\label{Sfree}
\eeq

\subsubsection{Current representation of the
asymptotically free model}

In this section we will present a description of the asymptotically
free Lifshitz NLSM in terms of the adjoint current, already introduced
in Section 2.
Since the Lifshitz model shares the quantum properties of the standard
2-d Lorentzian model, one might assume that these properties  are
inherited through a {\it detailed balance} condition.
The detailed balance action in the Lifshitz context is
basically the Lifshitz action in $D+1$ spacetime dimensions with $z=2$
where the potential of the theory is constructed by squaring the
equations of motion of the (euclideanized) Lorentz symmetric theory in
$D$ spacetime dimensions. This is evidently a property of the
free Lifshitz scalar as well as
the asymptotically free gauge theory in five dimensions constructed
by Ho\v{r}ava~\cite{HoravaYM}.
In Appendix A.2 we demonstrate how the detailed balance action of the
$4+1$-dimensional $z=2$ marginal scalar interaction inherits the
quantum properties (in the sense of the RG flow of couplings) of the
'parent' 4-dimensional Lorentzian marginal interaction.

A naive application of the detailed balance principle in the context of the
NLSM would require the squaring of the equations of motion\footnote{these
are easily derived by the introduction of a Lagrange multiplier field for
the unimodulus constraint in the action
and the subsequent elimination of the multiplier}
of the 2-d action defined in~(\ref{W2})
\beq
\frac{\delta W}{\delta \e} = \triangle\e - (\ebe)\; \e = 0
\eeq
The detailed balance potential would therefore correspond to a
marginal $z=2$ operator
\beq
\frac{\delta W}{\delta \e} \cdot \frac{\delta W}{\delta \e} &=&
\left[\triangle \e - (\ebe)\; \e \right]^{2} \nn
&=& \triangle \e \cdot \triangle \e
+ (\ebe)^2 \e^{~2} - 2 (\ebe)(\ebe) \nn
 &=& {\cal{O}}_1 - {\cal{O}}_2
\eeq
This term differs from the potential in the asymptotically free model
~(\ref{Sfree}) by the  term
\beq
\Delta L = 2 {\cal{O}}_2 - 2{\cal{O}}_3
\eeq
in the action density.
In $D = 2$ (only) this term can be rewritten with the help of the
antisymmetric tensor as
\beq
 \Delta L &=& 2 (\diedie) (\djedje) - 2 (\diedje) (\diedje) \nn
&=& 2 \; \epsilon_{ij} \epsilon_{kl} (\die \cdot \dke) (\dje \cdot \dle)
\label{dL}
\eeq
The current representation of the 2-d NLSM introduced in Section 2 elucidates
greatly the meaning of these operators.
Squaring equation~(\ref{dJ1}) we have
\beq
tr  \left[ (\partial \cdot {\bf J})^2 \right] = -2
\left[ \triangle \e \cdot \triangle \e - (\ebe)^2  \right]
= -2 {\cal{O}}_1 + 2 {\cal{O}}_2
\eeq
On the other hand, squaring the divergence of the dual current
${\bf \Jd_\mu} = \epsilon_{\mu \nu} {\bf J_\nu}$ (equation~\ref{dJ2}) we
get
\beq
tr  \left[ (\partial \cdot {\bf \Jd})^2 \right] &=& 4 \epsilon_{\mu \nu}
\epsilon_{\rho \sigma} \dmu e^a \dnu e^b \partial_\rho e^b \partial_\sigma
e^a
= - 4 \epsilon_{\mu \nu}\epsilon_{\rho \sigma}
(\dmu \vec{e}\cdot \partial_\rho \vec{e})
(\dnu \vec{e}\cdot \partial_\sigma \vec{e}) \nn
&=& -4 {\cal{O}}_2 + 4 {\cal{O}}_3
\eeq
Introducing also the temporal component of the adjoint current in the
$2+1$ dimensional  Lifshitz model
\be
J_t^{(a,b)} = e^a \dt e^b -e^b \dt e^a
\ee
we can express the asymptotically free action as
\be
S_{\mbox{\scriptsize asym.~free}} = - \frac{1}{4g^2} \int dt d^2 x \;
tr \Big[{\bf J_t J_t} -
(\partial \cdot {\bf J})^2 - (\partial \cdot {\bf \Jd})^2 \Big]
\ee
The analogy to the 2-d model can be made even closer through the
introduction of the complex adjoint vector
\be
{\bf Z_\mu} = {\bf J_\mu} + i {\bf \Jd_\mu} ~~~~~~~(\mu = 1, 2)
\ee
Due to the properties
\be
tr {\bf \Jd_\mu \Jd_\mu} = tr {\bf J_\mu J_\mu}
~~~~~,~~~~~~tr {\bf J_\mu \Jd_\mu} = 0 ~,
\ee
the 2-d model action~(\ref{W2}) is expressed as
\be
W[\e] = - \frac{1}{8g^2} \int d^2 x  \,tr \left [ {\bf J_{\mu} J_{\mu}}
+ {\bf \Jd_{\mu} \Jd_{\mu}} \right]
=  - \frac{1}{8g^2} \int d^2 x  \, tr~ {\bf Z} \cdot \bar{\bf Z}
\ee
while the asymptotically free Lifshitz model is compactly written
\beq
S_{\mbox{\scriptsize asym.~free}} = -\frac{1}{4 g^2} \int dt d^2 x \; tr
\Big[{\bf J_t J_t} -
(\partial \cdot {\bf Z})(\partial \cdot \bar{\bf Z}) \Big] ~.
\eeq

\section{Conclusions}

In this work we presented a study of the Lifshitz $O(N)$-symmetric NLSM
in 2+1 spacetime dimensions with a dynamical critical exponent $z=2$. The
general model includes three marginal dimension four operators with
three independent dimensionless couplings. The examination of the
one-loop contributions to the two-point function is instrumental in
identifying the one-coupling model which is asymptotically free
with the beta function in complete agreement to the conventional
NLSM in 1+1 dimensions.

Quantum inheritance is manifest therefore in the NLSM between two and
three dimensions although the action which inherits the asymptotic freedom
does not follow from the naive squaring of the 2-d equations of motion.
Instead, it admits an elegant
representation in terms of a complexified adjoint current which involves
both the (classically) conserved and its dual current in the two spatial
directions.

The known physics of the 2-d model are expected to appear
in the Lifshitz NLSM 'tuned' action.
The scale invariance of the 3-d action is broken dynamically by
quantum fluctuations and a scale will be introduced in the quantum
theory through the usual dimensional transmutation effect.
Excitations above the degenerate
vacuum are expected to become massive --as a result the $O(N)$ symmetry
will not break spontaneously in the ground state of the 3-d model.
This consists a violation of the Coleman-Mermin-Wagner theorem~\cite{Coleman}
which states that
long range order is not permitted in the ground state
of a theory with globally symmetric classical vacuum in two dimensions only.
The reason for this
violation is the anisotropic nature of the Lifshitz point.
A quick examination of the one-loop vacuum graph 
which determines   $\langle \sigma \rangle$ (Figure 2a) 
in $D+1$-dimensions
with anisotropy exponent $z$ shows that the logarithmic singularity
appears at $D=z$.
This is in accordance to the large-$N$
study of the $CP^{N-1}$ model~\cite{CPN} which established
asymptotic freedom and dynamical mass generation for all the models
in $D=z$ spatial dimensions.
We conclude therefore that the critical dimension for the lack of
long range order will be shifted in the Lifshitz point at $z+1$ dimensions.

An equally important observation is the lack of the
relevant dimension two operator in the quantum action of the
asymptotically free NLSM -- at least in the leading order.
It is feasible therefore that Lorentz symmetry
will not appear in the low energy regime of this theory, in contrast to
the generic expectation confirmed already in other models~\cite{LL,Yukawa}.
Instead, pions will propagate with a Galilean-type dispersion relation
$\omega^2 = p^4 + m^4$ in three dimensions.
Monte Carlo simulations of an
appropriate discretization of the Euclidean NLSM action should
be able to confirm such behavior in the single disordered phase of
this model.

The case of the Abelian rotor (or XY) model worked out
in Appendix B is consistent
with the above expectations.
At $N=2$ the 'tuned' action contains trivially a free
massless boson with an anisotropic dispersion relation  $\omega^2 = p^4$.
The interest here lies in the examination of  the
order of the transition between the
massless non-relativistic phase and the disordered phase for the
lattice Lifshitz action.

\vspace{1cm}

\nin{\bf Acknowledgments}
This work is partly supported by the National Technical University of
Athens through the Basic Research Support Programme 2008.

\appendix

\begin{center}

{\Large \bf Appendix}

\end{center}

\section{Scalar field theory at the Lifshitz point}

\subsection{Renormalizability and power counting}

A scalar field theory at the Lifshitz point is constructed by considering the
fixed point with anisotropic scaling between time and $D$-spatial dimensions.
Assuming a dynamical critical exponent $z$ which governs the anisotropy
\be
t \rightarrow b^z\, t \;\;\; ,  \;\;\;
x_i \rightarrow b \,x_i \;\;\;\;(i = 1,\dots ,D)  \;\;,
\ee
the free fixed point action is constructed ($\Delta = \partial_i \partial_i$ is the Euclidean Laplacian)
\be\label{scalaraction}
S_b=\frac{1}{2}\int dt d^Dx \left( \dot\phi^2-\phi \left(-\Delta\right)^{z}\phi \right).
\ee
Canonical dimensions are assigned to fields and spacetime arguments as
\be
[x_\kappa]=-1~~~~~~~~[t]=-z~~~~~~~~[\phi]=\frac{D-z}{2}.
\ee
Plane waves propagate in the theory~(\ref{scalaraction}) with dispersion
relation
\be
\omega=(\textbf{p}^2)^\frac{z}{2}.
\ee
Quantization of the theory is straightforward --a mass term can also
be considered by adding $-\frac{1}{2}m_b^{2z} \phi^2$ to the
action~(\ref{scalaraction}) where $[m_b]=1$.
The scalar field propagator is then written
\be\label{Gbm}
G_b(\omega, \textbf{p})=\frac{i}{\omega^2-(\textbf{p}^2)^z-m_b^{2z}
+i\varepsilon}
\ee
Interacting theories are constructed by the addition of non-gaussian terms
to~(\ref{scalaraction}). Perturbative renormalizability is possible and
examined through standard power counting arguments.
Polynomial interactions of the type $\lambda \phi ^n$ are marginal if
$[\lambda]=0$ i.e.
for a critical power
\be
n_{cr} = \frac{2(D+z)}{D-z}
\ee
and relevant for $n <n_{cr}$. In particular, the scalar theory
at $D=z$ will be power counting renormalizable to all orders in
perturbation theory.

Furthermore, marginal interactions of the $\lambda (\di \phi)^2 \phi^n$
type are also allowed now for $z < D$ since
positive integer values are possible for
\be
n = \frac{4(z-1)}{D-z}
\ee
It can also be checked that marginal interactions of the type
\be
\lambda (\di \phi)^2 (\partial_j \phi)^2 \phi^n
\ee
with $n \ge 0$
will not appear in the $z=2$, $D \ge 3$ theory and are possible for $z=3$
((D=4, n=2) or (D=5, n=0)) or higher values of $z$.

\subsection{$z=2$ and detailed balance}

In the following we examine the relation between the so-called
'detailed balance' scalar action (which is a particular $z=2$ action) in $D+1$
dimensions and the Lorentz symmetric theory in $D-$ (Euclideanized) spacetime
dimensions. We will demonstrate that in fact the detailed balance theory at
the Lifshitz point shares the same quantum properties with the 'parent'
Lorentzian theory, in the sense that the marginal couplings of both theories
run with the same beta function.

As a specific example we examine the $z=2$ scalar theory in $4+1$ dimensions.
The general action at the UV fixed point including the marginal couplings is
written
\be
\label{Sz2}
S_{z=2}= \int dt d^4 x \left( \hf \dot\phi^2 - \hf \left( \Delta \phi \right)^2 -
\lambda_1 \phi^6 -
\lambda_2 (\di \phi)^2 \phi^2 \right)
\ee
where $\Delta = \di \di$ denotes the Euclidean Laplacian in $D=4$.
This theory has been examined in detail in~\cite{iengo} where the
one loop beta functions for the running of the couplings $\lambda_1$ and
$\lambda_2$ have been calculated in dimensional regularization.
The results --after the rescaling of equations~(3.9) in~\cite{iengo} --
are\footnote{redefine the marginal couplings in~\cite{iengo} as
$\lambda_1 = \kappa/6!$ and $\lambda_2 = g/4$.}
\beq
\label{beta1}
\beta_{\lambda_1} &=& \frac{d \lambda_1}{d \ln \mu} =
\frac{15}{16 \pi^2}\lambda_1 \lambda_2 - \frac{\lambda_2^3}{64 \pi^2} \\
\label{beta2}
\beta_{\lambda_2} &=& \frac{d \lambda_2}{d \ln \mu} =
\frac{3}{16 \pi^2} \lambda_2^2
\eeq
As demonstrated in~\cite{iengo} both couplings are IR free.

On the other hand, the detailed balance action
in $D+1$ spacetime dimensions with $z=2$ is constructed by
a standard kinetic term in the time direction and a potential term which
is the square of the
equations of motion of a $D$-dimensional Euclidean theory
\be
S_{\mbox{\scriptsize det.bal.}}=\frac{1}{2}\int dt d^Dx \left( \dot\phi^2
-\frac{1}{\kappa^2} \left( \frac{\delta W[\phi]}{\delta\phi}\right)^2 \right),
\ee
where $W[\phi]$ is the Euclidean action of a relativistic scalar in $D$
dimensions and the dimensionless parameter $\kappa$ can be absorbed by a
rescaling of the time variable. Notice that the canonical
dimension of $\phi$ remains unchanged in the relativistic theory in
$D-$dimensions and in the $D+1$ theory with $z=2$.

The marginal part of the $D=4$ Euclidean theory is
\be W[\phi] =
\int d^4x \left( \hf \di \phi \di \phi + \frac{g}{4!} \phi^4
\right).
\ee
from which we obtain the potential term
\be \left(
\frac{\delta W[\phi]}{\delta\phi}\right)^2 = \left(-\Delta \phi +
\frac{g}{6} \phi^3\right)^2 = \left(\Delta \phi \right)^2 +
\frac{g^2}{36} \phi^6 - \frac{g}{3}\phi^3 \Delta \phi
\ee
Integrating by parts we can reexpress
\be \phi^3 \Delta \phi = -3
\phi^2 (\di \phi)^2 + {\rm total \; derivative}
\ee
from which the
couplings in eq.~(\ref{Sz2}) are identified as following in the
detailed balance action
\be \label{detbal} \lambda_1 =
\frac{g^2}{36} \hspace{1.5cm} \lambda_2 = g \;.
\ee
It is
recognized now that the coupling $\lambda_2$ runs with the
standard beta function of the $D=4$ relativistic $\phi^4$ theory,
eq.~({\ref{beta2}). Furthermore, at the detailed balance point
$\lambda_1 = g^2/36$, eq.~({\ref{beta1}) becomes \be
\beta_{\lambda_1} = \frac{2 g}{36}\beta_{g} = \frac{15}{16
\pi^2}\frac{g^3}{36} - \frac{g^3}{64 \pi^2} \ee which is satisfied
precisely by \be \label{beta4} \beta_{g} = \frac{3}{16 \pi^2} g^2
. \ee It is understood therefore that the detailed balance action
preserves the precise constraint of the marginal
couplings~(\ref{detbal}) under the renormalization group (RG)
flow. In fact the RG flow is
controlled by the flow of the lower dimensional marginal coupling
$g$ and in that sense the $4+1$ theory at the $z=2$ Lifshitz point
{\it inherits the quantum mechanical properties of the lower
dimensional relativistic theory}. Seen differently, a deviation
from the detailed balance point
\be \lambda_1 = \frac{g^2}{36} +
\delta \hspace{1.5cm} \lambda_2 = g \;. \label{gd}
\ee
will produce an RG running for $\delta$. Substituting~(\ref{gd}) in
eq.~({\ref{beta1}) determines
\be \label{betad} \beta_{\delta} =
\frac{d \;\delta}{d\ln \mu}= \frac{15}{16 \pi^2}{g}\;\delta
\ee
from which it is confirmed that the detailed balance point $\delta
= 0$ is a fixed point of the RG flow. It is interesting to note
also that since $g$ runs slowly to the IR fixed point --in fact
logarithmically due to~(\ref{beta4}) --, $\delta$ also runs
slowly, to the IR free fixed point as seen by
integrating~(\ref{betad})
\be \frac{\delta}{\delta_o} =
\left(\frac{g}{g_o}\right)^5
\ee
where $\delta_o$, $g_o$ are fixed
values of the couplings at some arbitrary energy scale. We
conclude therefore that the relative deviation of the $\lambda_1$
coupling from the detailed balance point will also diminish slowly
as
\be \frac{\delta}{\lambda_1}\sim\frac{\delta}{g^2} \sim g^3
\ee
and the detailed balance point will {\it attract the RG flow} in
the deep IR regime.

\section{The Abelian Rotor at the Lifshitz point}

The case of the $XY$ or quantum rotor model $(N=2)$ can be studied by the
introduction of a compact field $\theta(x)$ such as
\be
\e = (\cos\theta, \sin\theta)
\ee
The relativistic model is simply a free massless boson
\be
W^{XY}[\e] = \frac{1}{2g^2} \int d^2 x \; \dithh
\ee
Using the derivatives
\beq
\die &=& (-\sin\theta, \cos\theta)\;\dith \\
\Delta\e &=& (-\sin\theta, \cos\theta)\;\boxth - \dithh\;\e
\eeq
the marginal operators at the $z=2$ Lifshitz point take the form
\beq
{\cal{O}}_1 &=&  (\boxth)^2 + (\dithh)^2 \nn
{\cal{O}}_2 &=& {\cal{O}}_3 = (\dithh)^2
\eeq
It is seen therefore that the 'tuned action' is the one that
inherits the free massless property at the Lifshitz point since the
four-field term cancels out and the
action~(\ref{Sfree}) becomes simply
\be
S^{XY} = \frac{1}{2g^2} \int dt d^2 x \Big[
(\dtth)^2 - (\boxth)^2
\Big]
\label{XY}
\ee
Beyond the 'tuned point' the incorporation of the
four-field interaction term
$(\dithh)^2$ in the model would affect non-trivially
the dynamics.

It would also be very interesting to study the nature of the transition
of the lattice regularized action~(\ref{XY}) for the quantum rotor from the
high temperature phase to the long range ordered phase corresponding to
a free Lifshitz scalar with dispersion relation
\be
\omega^2 = k^4
\ee
It is well known that the 2-d relativistic model possesses a
Kosterlitz-Thouless (KT) type of phase transition and it will be interesting
to check if the infinite order of the KT transition will be inherited to
the lattice Lifshitz model.

\end{document}